\documentclass[10pt,twocolumn,amssymb,amsmath,aps,pra]{revtex4-1}
\usepackage{graphicx}

\begin{document}

\title{Light-driven mass density wave dynamics in optical fibers}
\date{August 10, 2018}
\author{Mikko Partanen}
\author{Jukka Tulkki}
\affiliation{Engineered Nanosystems group, School of Science, Aalto University, P.O. Box 12200, 00076 Aalto, Finland}

\begin{abstract}
We have recently developed the mass-polariton (MP) theory of light to describe the light propagation in
transparent bulk materials [Phys.~Rev.~A \textbf{95}, 063850 (2017)].
The MP theory is general as it is based on the covariance principle and the fundamental conservation laws of nature. Therefore, it can be applied also to nonhomogeneous and dispersive materials.
In this work, we apply the MP theory of light to describe propagation of light in step-index
circular waveguides. We study the eigenmodes of the electric and magnetic fields in a
waveguide and use these modes to calculate the optical force density, which is used
in the optoelastic continuum dynamics (OCD) to simulate the dynamics of medium atoms in the waveguide.
We show that the total momentum and angular momentum in the waveguide
are carried by a coupled state of the field and the medium. In particular,
we focus in the dynamics of atoms, which has not been covered in previous theories
that consider only field dynamics in waveguides.
We also study the elastic waves generated in the waveguide
during the relaxation following from atomic displacements in the waveguide.
\end{abstract}

\maketitle

\onecolumngrid
\vspace{-0.2cm}
\twocolumngrid

\section{Introduction}

The linear and angular momenta of light have been recently investigated in vacuum \cite{Bliokh2013a,Bliokh2014b},
in various photonic materials \cite{Bliokh2015b,Du2017,Alpeggiani2018,Bekshaev2015},
in the quantum domain \cite{Mair2001,MolinaTerriza2007},
and also in the near-field regime \cite{Lin2013,Bliokh2017a,Bliokh2017b,Bliokh2014a}.
Part of the recent progress has been
covered in books \cite{Andrews2013,Torres2011} and reviews
\cite{Milonni2010,Yao2011,Piccirillo2013,Dressel2015,Bliokh2015a}.
However, despite the long history of the studies of optical waveguides \cite{Kapany1972,Snyder1983},
the detailed theoretical description of the linear and angular momenta of light in waveguides has been missing.
In particular, the coupled dynamics of the field and the medium atoms
and its relation to the linear and angular momenta of light have been considered recently for homogeneous
dielectric materials \cite{Partanen2017c,Partanen2017e,Partanen2018a},
but detailed studies of the effects arising from the coupled field-medium dynamics do not exist for waveguides.
The works on the coupled dynamics also question the conventional approximation of fixed atomic positions
in the description of propagation of light in a medium by showing that the atomic
dynamics is both necessary for the fulfillment of Newton's first law and it also carries
an essential well-defined part of the total linear and angular momenta of light in a medium.
In the present work, we study the coupled field-medium dynamics
and related effects beyond the conventional approximation of fixed atoms
in waveguides.

Combining the electrodynamics of continuous media, elasticity theory, and Newtonian mechanics,
it has been shown that a light pulse propagating in
a medium drives forwards an atomic mass density wave
(MDW), which carries a major part of the total linear and angular momenta
of light in many common dielectrics \cite{Partanen2017c,Partanen2017e,Partanen2018a}.
In the classical optoelastic continuum dynamics (OCD), formulated by us, the MDW
is an inevitable consequence of the optical
force density of the electromagnetic field.
When the optical force density and the elastic forces are included in the Newtonian
equation of motion, one can solve numerically the space-time dynamics of the medium.
By considering the numerically calculated space-time dynamics of the
atoms in the medium, one can see that field intensity maxima are accompanied
by atomic density maxima, which propagate
with the same velocity as the field intensity maxima.
Note, in particular, that these MDWs should not be confused
with elastic waves, the dynamics of which is governed by elastic forces
and which propagate with the velocity of sound.
Instead, the MDWs should sooner be considered as shock waves,
which are generated by the optical force field.

In the single-photon picture, the coupling of the electromagnetic
field to the atomic MDW gives rise to mass-polariton (MP)
quasiparticles, which are covariant coupled states of the field
and matter having a nonzero rest mass \cite{Partanen2017c,Partanen2017e}.
In this MP quasiparticle model, the MDW follows directly
from the Lorentz transformation of the energy and momentum of light.
Hence, the MP is fundamentally
different from the conventional exciton-polariton and the
phonon-polariton quasiparticles, in which a photon is in
resonance or in close resonance with an internal excited state of the medium.
Within the framework of the MP theory of light, the correct
way of thinking the exciton-polariton and the phonon-polariton quasiparticles
is to think them as coupled states of the MP and an exciton or an optical phonon, respectively.
Note, however, that in the present work we limit our investigations to
highly transparent materials in which the coupling of the MPs to
the excitons or optical phonons is very weak, and therefore, we do not
elaborate this coupling further in this work.

The MP quasiparticles have been shown to carry a total momentum
of the Minkowski form $p_\mathrm{MP}=n_\mathrm{p}\hbar\omega/c$, where
$n_\mathrm{p}$ is the phase refractive index of the medium,
$\hbar$ is the reduced Planck constant, $\omega$ is the angular
frequency, and $c$ is the speed of light in vacuum \cite{Partanen2017e}.
The total MP momentum is split between the electromagnetic field
and the MDW so that the share of the field corresponds to
the Abraham momentum $p_\mathrm{field}=\hbar\omega/(n_\mathrm{g}c)$,
where $n_\mathrm{g}$ is the group refractive index.
The MDW carries the difference of the Minkowski and Abraham momenta.
Therefore, the MP theory of light provides a resolution
to the centennial Abraham-Minkowski controversy of photon momentum in a medium
\cite{Pfeifer2007,Barnett2010a,Barnett2010b,Bliokh2017a,Bliokh2017b,Leonhardt2014,Brevik2017}.
Both momenta are right when they are defined in the right context.

In this work, we apply the MP theory to simulate the propagation of 
the MDW of selected light pulses in a step-index circular waveguide.
It is well-known that, in waveguides, part of the electromagnetic field propagates
as an evanescent field in the cladding layer surrounding the core.
In the MP theory of light, this affects the atomic MDW that propagates
partly in the core and partly in the cladding layer. The total magnitude of the
the MDW is, however, fully determined by the covariant state of light
in a medium \cite{Partanen2017c,Partanen2017e}.
If the cladding layer is vacuum,
the entire MDW is propagating in the core material
and its magnitude should be reduced when compared to the MDW in the case of a homogeneous dielectric.
This reduction of the MDW should also be related to the waveguide dispersion.
Here we show that the numerical OCD simulations accurately verify these points and
the full correspondence with the MP quasiparticle model. This also
shows that the optical force density used in the OCD model in homogeneous
materials also accurately applies to waveguides.

Note that there exist some previous works that address the question of consistent definition of the angular
momentum of light in circular waveguides \cite{Kien2006,Gregg2015,Alexeyev1998}.
These works do not, however, consider the essential role of the atomic MDW
as a carrier of the major part of the total angular momentum of light
in waveguides. Therefore, these works have inherent limitations like (1) not being able
to predict the transfer of atomic density with light, and
(2) not describing the inevitable
dissipation of field energy resulting from the displacement of waveguide atoms.
In contrast, these aspects are directly obtained in the present work from the numerical OCD simulations.

This work is organized as follows: Section \ref{sec:theory} reviews
the theoretical foundations of the OCD model.
Section \ref{sec:fields} presents the solution of the electric
and magnetic fields of a light pulse in a step-index circular waveguide geometry.
The OCD simulations of the propagation of selected light pulses
in a circular waveguide are presented in Sec.~\ref{sec:simulations},
where we also describe the elastic relaxation dynamics of the nonequilibrium
atomic displacements that the MDW leaves in the waveguide.
In Sec.~\ref{sec:comparison},
we compare the OCD results with the results obtained
by using the MP quasiparticle model.
Finally, conclusions are drawn in Sec.~\ref{sec:conclusions}.

\section{\label{sec:theory}Optoelastic continuum dynamics}

\subsection{Beyond the approximation of fixed atoms}

In the approximation of fixed atoms, the atomic nuclei are bound to
their fixed equilibrium positions and atoms respond to the electromagnetic
field only through their polarization. This approximation is deeply rooted
in the foundations of conventional electrodynamics of continuous media \cite{Landau1984,Jackson1999}.
The atoms have been traditionally assumed to be fixed in a medium as their
total mass energy is extremely large in comparison with typical
energies of optical fields. Therefore, in short time scales of optical
fields, the atoms that interact with photons can at most move
by an exceedingly small amount during the interaction. Conventionally, this
small atomic movement has been thought to be fully negligible.
Note that, if the atoms were fixed in a solid, there would not be elastic or sound waves
either because these waves are made from displacement of atoms around
their equilibrium positions. If the elastic forces can displace atoms
around their equilibrium positions, why could not the optical force desplace them?

Very recently, the results of computer simulations using
the OCD model \cite{Partanen2017c,Partanen2017e,Partanen2018a},
which couples the electrodynamics of continuous media to the elasticity
theory and Newtonian mechanics of atoms, have questioned
the approximation of fixed atoms by showing that the collective small motion
of atoms forms an atomic MDW. As coupled systems have long been under detailed investigations
in many fields of physics, it is surprising that the coupled dynamics of light and matter has not
been studied in detail already much earlier.

\subsection{Newton's equation of motion}

Light is optoelastically coupled to the medium by Newton's
equation of motion. This is one of the key elements of the OCD model.
Since the force density of the optical field can give only small
velocities for the atoms of a continuous medium, the atomic
velocities are certainly nonrelativistic.
Therefore, Newton's equation of motion for the mass density
of the medium $\rho_\mathrm{a}(\mathbf{r},t)$,
which is equal to the number density of atoms times the atomic mass, can be written as
\begin{equation}
 \rho_\mathrm{a}(\mathbf{r},t)\frac{d^2\mathbf{r}_\mathrm{a}(\mathbf{r},t)}{dt^2}=\mathbf{f}_\mathrm{opt}(\mathbf{r},t)+\mathbf{f}_\mathrm{el}(\mathbf{r},t),
 \label{eq:mediumnewton}
\end{equation}
where $\mathbf{r}_\mathrm{a}(\mathbf{r},t)$ is the
instantaneous position- and time-dependent atomic
displacement field of the medium, $\mathbf{f}_\mathrm{opt}(\mathbf{r},t)$
is the optical force density experienced by the medium atoms,
and $\mathbf{f}_\mathrm{el}(\mathbf{r},t)$ is the elastic force
density, which arises between atoms that are displaced from
their initial equilibrium positions by the optical force density.

In previous literature, several forms of the optical force
density have been extensively discussed \cite{Milonni2010,Brevik1979}.
We have recently shown, for a low-loss dielectric, that there is only
one form of optical force density that is fully consistent
with the MP quasiparticle model and the underlying principles
of the special theory of relativity \cite{Partanen2017c}.
This force density is presented for a nondispersive
medium with a refractive index $n$ as \cite{Milonni2010,Partanen2017c}
\begin{equation}
 \mathbf{f}_\mathrm{opt}(\mathbf{r},t)=-\frac{\varepsilon_0}{2}\mathbf{E}^2\nabla n^2+\frac{n^2-1}{c^2}\frac{\partial}{\partial t}\mathbf{E}\times\mathbf{H}.
 \label{eq:opticalforcedensity}
\end{equation}
This optical force density leads to the transfer of an inevitable
part of the total momentum of light by the MDW \cite{Partanen2017c,Partanen2017e}.
In this work, we show that the optical force density in
Eq.~\eqref{eq:opticalforcedensity} is
also applicable in the description of optical forces
of confined modes, such as light propagating in
optical fibers or waveguides.

In the OCD simulations, we use the well-known elastic force density that follows from
Hooke's law \cite{Kittel2005}. Using the material displacement
field $\mathbf{r}_\mathrm{a}(\mathbf{r},t)$, this elastic force density can be given
for an isotropic linear elastic medium as \cite{Bedford1994}
\begin{equation}
 \mathbf{f}_\mathrm{el}(\mathbf{r},t)=\textstyle(\lambda_\mathrm{L}+2\mu_\mathrm{L})\nabla[\nabla\cdot\mathbf{r}_\mathrm{a}(\mathbf{r},t)]-\mu_\mathrm{L}\nabla\times[\nabla\times\mathbf{r}_\mathrm{a}(\mathbf{r},t)],
 \label{eq:elasticforcedensity}
\end{equation}
where $\lambda_\mathrm{L}$ and $\mu_\mathrm{L}$ are the Lam\'e elastic constants of the medium.
The Lam\'e elastic constants could also be replaced by any two independent
elastic moduli, such as the bulk and shear moduli whose
relation to the Lam\'e constants is described in \cite{Mavko2003}.
For anisotropic cubic crystals, such as silicon, the elastic force density
has a slightly more complicated form, which
is given, e.g., in \cite{Partanen2017e,Kittel2005}.

When we apply the OCD model to simulate the propagation
of light pulses in optical waveguides in Sec.~\ref{sec:simulations},
we use a perturbative approach in which we neglect
the extremely small damping of the field that results from
the transfer of field energy to the kinetic and elastic energies
of the medium atoms by the optical force density in Eq.~\eqref{eq:opticalforcedensity}.
In the case of light propagation in a homogeneous medium,
the accuracy of this approximation is estimated in \cite{Partanen2017c},
but the conclusions are also valid for waveguides provided
that optical losses are small, which is typically a good approximation
as light can travel in waveguides over long distances.

\subsection{Energies, momenta, and angular momenta of the field and the mass density wave}

The total electromagnetic energy of the light pulse is obtained as an integral
of the instantaneous energy density of the field and the total mass energy of the MDW is
obtained as an integral of the MDW mass density just as in the case of a homogeneous
medium \cite{Partanen2017c}. These relations and the expression of the total
MP energy of a light pulse are given by
\begin{align}
 E_\mathrm{MP}=\int\Big[\rho_\mathrm{MDW}c^2+\frac{1}{2}(\varepsilon\mathbf{E}^2+\mu\mathbf{H}^2)\Big]d^3r,\hspace{1.2cm}\nonumber\\[3pt]
 E_\mathrm{MDW}=\int\rho_\mathrm{MDW}c^2d^3r,\hspace{0.3cm}
 E_\mathrm{field}=\int\frac{1}{2}(\varepsilon\mathbf{E}^2+\mu\mathbf{H}^2)d^3r.
\label{eq:ocdenergy}
\end{align}
As shown in \cite{Partanen2017c}, the kinetic energy of the MDW is extremely small
and neglected in $E_\mathrm{MDW}$, which only describes the mass energy of the MDW.

Again, in the same way as in a homogeneous medium, the total momentum
of the coupled MP state of the field and matter and the momentum shares
of the electromagnetic field and the atomic MDW are given by \cite{Partanen2017c}
\begin{align}
 \mathbf{P}_\mathrm{MP}=\int\Big(\rho_\mathrm{a}\mathbf{v}_\mathrm{a}+\frac{\mathbf{E}\times\mathbf{H}}{c^2}\Big)d^3r,\hspace{1.2cm}\nonumber\\[3pt]
 \mathbf{P}_\mathrm{MDW}=\int\rho_\mathrm{a}\mathbf{v}_\mathrm{a}d^3r,\hspace{0.3cm}
 \mathbf{P}_\mathrm{field}=\int\frac{\mathbf{E}\times\mathbf{H}}{c^2}d^3r.
\label{eq:ocdmomentum}
\end{align}
In the MP theory of light, we also obtain the total angular momentum of the MP,
denoted by $\mathbf{J}_\mathrm{MP}$,
and its shares $\mathbf{J}_\mathrm{field}$ and $\mathbf{J}_\mathrm{MDW}$
carried by the electromagnetic field and the atomic MDW.
These angular momenta are given by \cite{Partanen2018a}
\begin{align}
 \mathbf{J}_\mathrm{MP}=\int\mathbf{r}\times\Big(\rho_\mathrm{a}\mathbf{v}_\mathrm{a}+\frac{\mathbf{E}\times\mathbf{H}}{c^2}\Big)d^3r,\hspace{1.6cm}\nonumber\\[3pt]
 \mathbf{J}_\mathrm{MDW}=\int\mathbf{r}\times\rho_\mathrm{a}\mathbf{v}_\mathrm{a}d^3r,\hspace{0.3cm}
 \mathbf{J}_\mathrm{field}=\int\mathbf{r}\times\Big(\frac{\mathbf{E}\times\mathbf{H}}{c^2}\Big)d^3r.
\label{eq:ocdangularmomentum}
\end{align}

\section{\label{sec:fields}Electric and magnetic fields of a light pulse in a circular waveguide}

\begin{figure}[b]
\centering
 \includegraphics[width=0.4\textwidth]{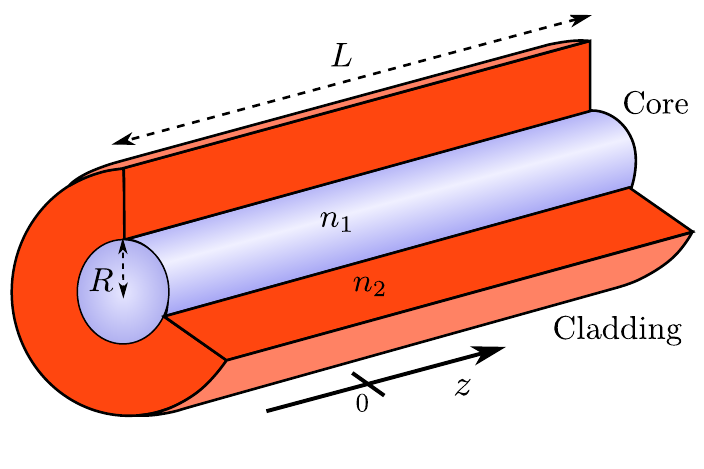}
\caption{\label{fig:waveguide}
Illustration of the step-index circular waveguide geometry.
The waveguide core with radius $R$ has a refractive index $n_1$
while the refractive index of the cladding layer is $n_2$. The total length
of the waveguide is $L$.}
\end{figure}

We aim at using the electric and magnetic fields of a light pulse in a waveguide
to simulate the propagation of the MDW driven by the light pulse.
Therefore, we next review the solution of the electric and magnetic fields of a light pulse
in a step-index circular waveguide geometry illustrated in Fig.~\ref{fig:waveguide}.
The waveguide core radius is $R$. Inside the core, the refractive index
is $n_1$ while the refractive index in the cladding layer is $n_2$.

\subsection{Fields in a step-index circular waveguide}

The electric and magnetic field modes are given as exact solutions of the Maxwell's equations
for the step-index circular waveguide geometry as presented, e.g., in \cite{Yariv2007}.
These solutions are called Bessel modes.
The necessary condition for the existence of confined modes is $n_1>n_2$.
The cladding layer is assumed to be thick enough so that the field of confined modes
is virtually zero in the outermost boundary of the cladding layer and we only
need to account for the core and the cladding when computing the spatial
distribution of the electromagnetic field inside the waveguide.

Here, we assume a light pulse with a nonzero spectral width as described by the
Gaussian function $u(k_0')=e^{-[(k_0'-k_0)/\Delta k_0]^2/2}/(\sqrt{2\pi}\Delta k_0)$,
where $k_0=\omega_0/c$ is the wavenumber in vacuum for central frequency $\omega_0$,
and $\Delta k_0$ is the standard deviation of the wavenumber in vacuum.
In the cylindrical coordinate system with coordinates $r$, $\phi$, and $z$
and basis vectors $\hat{\mathbf{r}}$, $\hat{\mathbf{\boldsymbol{\phi}}}$, and $\hat{\mathbf{z}}$,
the instantaneous electric and magnetic fields of the Bessel mode pulses
are given, in the waveguide core ($r<R$), as
\begin{align}
& \mathbf{E}_{\{r<R\}}(\mathbf{r},t)\nonumber\\
&=\mathrm{Re}\Big(\int_{-\infty}^\infty\Big\{
\frac{ik_z'}{h'^2}\Big[Ah'J_l'(h'r)+\frac{il\omega'\mu_0}{k_z'r}BJ_l(h'r)\Big]\hat{\mathbf{r}}\nonumber\\
&\hspace{0.5cm}+\frac{ik_z'}{h'^2}\Big[\frac{il}{r}AJ_l(h'r)-\frac{\omega'\mu_0}{k_z'} Bh'J_l'(h'r)\Big]\hat{\boldsymbol{\phi}}\nonumber\\
&\hspace{0.5cm}+AJ_l(h'r)\hat{\mathbf{z}}\Big\}
e^{il\phi}u(k_0')e^{i[k_z'z-\omega(k_z')t]}dk_0'\Big),
\label{eq:efield1}
\end{align}
\begin{align}
& \mathbf{H}_{\{r<R\}}(\mathbf{r},t)\nonumber\\
&=\mathrm{Re}\Big(\int_{-\infty}^\infty\Big\{
\frac{ik_z'}{h'^2}\Big[Bh'J_l'(h'r)-\frac{il\omega'\varepsilon_1'}{k_z'r}AJ_l(h'r)\Big]\hat{\mathbf{r}}\nonumber\\
&\hspace{0.5cm}+\frac{ik_z'}{h'^2}\Big[\frac{il}{r}BJ_l(h'r)+\frac{\omega'\varepsilon_1'}{k_z'} Ah'J_l'(h'r)\Big]\hat{\boldsymbol{\phi}}\nonumber\\
&\hspace{0.5cm}+BJ_l(h'r)\hat{\mathbf{z}}\Big\}
e^{il\phi}u(k_0')e^{i[k_z'z-\omega(k_z')t]}dk_0'\Big),
\label{eq:hfield1}
\end{align}
and, in the cladding layer ($r>R$), as
\begin{align}
& \mathbf{E}_{\{r>R\}}(\mathbf{r},t)\nonumber\\
&=\mathrm{Re}\Big(\int_{-\infty}^\infty\Big\{
-\frac{ik_z'}{q'^2}\Big[Cq'K_l'(q'r)+\frac{il\omega'\mu_0}{k_z'r}DK_l(q'r)\Big]\hat{\mathbf{r}}\nonumber\\
&\hspace{0.5cm}-\frac{ik_z'}{q'^2}\Big[\frac{il}{r}CK_l(q'r)-\frac{\omega'\mu_0}{k_z'} Dq'K_l'(q'r)\Big]\hat{\boldsymbol{\phi}}\nonumber\\
&\hspace{0.5cm}+CK_l(q'r)\hat{\mathbf{z}}\Big\}
e^{il\phi}u(k_0')e^{i[k_z'z-\omega(k_z')t]}dk_0'\Big),
\label{eq:efield2}
\end{align}
\begin{align}
& \mathbf{H}_{\{r>R\}}(\mathbf{r},t)\nonumber\\
&=\mathrm{Re}\Big(\int_{-\infty}^\infty\Big\{
-\frac{ik_z'}{q'^2}\Big[DqK_l'(q'r)-\frac{il\omega'\varepsilon_2}{k_z'r}CK_l(q'r)\Big]\hat{\mathbf{r}}\nonumber\\
&\hspace{0.5cm}-\frac{ik_z'}{q'^2}\Big[\frac{il}{r}DK_l(q'r)+\frac{\omega'\varepsilon_2'}{k_z'} Cq'K_l'(q'r)\Big]\hat{\boldsymbol{\phi}}\nonumber\\
&\hspace{0.5cm}+DK_l(q'r)\hat{\mathbf{z}}\Big\}
e^{il\phi}u(k_0')e^{i[k_z'z-\omega(k_z')t]}dk_0'\Big).
\label{eq:hfield2}
\end{align}
Here $J_l(x)$ are the Bessel functions of the first kind,
$K_l(x)$ are the modified Bessel functions of the second kind,
$h=\sqrt{k_1^2-k_z^2}$, $q=\sqrt{k_z^2-k_2^2}$, and $\omega(k_z')=ck_z'/n_\mathrm{p,eff}'$
is the waveguide dispersion relation with the effective phase index denoted by $n_\mathrm{p,eff}$
and the wavenumbers of the core and cladding are given by $k_1=n_1k_0$ and $k_2=n_2k_0$, respectively.
The derivatives of the Bessel functions are denoted by $J_l'(x)$ and $K_l'(x)$.
The other primed quantities in Eqs.~\eqref{eq:efield1}--\eqref{eq:hfield2}
correspond to the values obtained for $k_0'$ instead of $k_0$.
The materials are here assumed to be nondispersive.
The unknown amplitude factors $A$, $B$, $C$, and $D$ can be related to each other by applying
the boundary conditions at the interface between the core and the cladding as described below
or, e.g., in \cite{Yariv2007}.

In this work, we apply the fields above to describe the propagation of light pulses with a relatively small spectral width.
In this limit, the light pulse is approximately monochromatic and
Eqs.~\eqref{eq:efield1}--\eqref{eq:hfield2} can be approximated further
in the same way as presented for Laguerre-Gaussian pulses
in a homogeneous medium in \cite{Partanen2018a}.

\subsection{Characteristic equation and the relations between the field amplitudes}

The unknown amplitude factors $A$, $B$, $C$, and $D$ can be related to each other by
the boundary conditions. Requiring that $E_\phi$, $E_z$, $H_\phi$, and $H_z$ are continuous  at the
interface between the core and the cladding at $r=R$ leads to four equations, given by
\begin{equation}
 AJ_l(hR)=CK_l(qR),
 \label{eq:bc1}
\end{equation}
\begin{equation}
 BJ_l(hR)=DK_l(qR),
 \label{eq:bc2}
\end{equation}
\begin{equation}
A\frac{ilk_z}{\omega\mu_0R}\Big(\frac{1}{q^2}+\frac{1}{h^2}\Big)=B\Big[\frac{J_l'(hR)}{hJ_l(hR)}+\frac{K_l'(qR)}{qK_l(qR)}\Big],
 \label{eq:bc3}
\end{equation}
\begin{equation}
A\Big[\frac{n_1^2J_l'(hR)}{hJ_l(hR)}+\frac{n_2^2K_l'(qR)}{qK_l(qR)}\Big]=B\frac{lk_z}{i\omega\varepsilon_0R}\Big(\frac{1}{q^2}+\frac{1}{h^2}\Big).
 \label{eq:bc4}
\end{equation}

Equations \eqref{eq:bc3} and \eqref{eq:bc4} lead to the characteristic equation for $k_z$ from which
the coefficients $A$ and $B$ have been cancelled. This characteristic equation is transcendental and it reads
\begin{align}
&\Big(\frac{lk_z}{k_0R}\Big)^2\Big(\frac{1}{q^2}+\frac{1}{h^2}\Big)^2\nonumber\\
&=\Big[\frac{J_l'(hR)}{hJ_l(hR)}+\frac{K_l'(qR)}{qK_l(qR)}\Big]
\Big[\frac{n_1^2J_l'(hR)}{hJ_l(hR)}+\frac{n_2^2K_l'(qR)}{qK_l(qR)}\Big].
\label{eq:characteristic}
\end{align}
For a given $l$ and a given frequency $\omega$, there will be only a finite number
of solutions $k_z$ that satisfy Eq.~\eqref{eq:characteristic}. Once a specific solution is known,
one can use Eqs.~\eqref{eq:bc1}--\eqref{eq:bc4} to solve the ratios of the amplitude factors $A$, $B$, $C$, and $D$.
The absolute magnitudes of the amplitude factors become determined independent
of a free phase factor by normalizing the fields
so that the total electromagnetic energy of the light pulse is $E_\mathrm{field}$.

\subsection{Waveguide dispersion}

The effective phase refractive index $n_\mathrm{p,eff}$ for the field propagating
in the optical fiber is given as the ratio of $k_z$,
solved from the characteristic equation in Eq.~\eqref{eq:characteristic},
and $k_0$ as \cite{Yariv2007}
\begin{equation}
 n_\mathrm{p,eff}=\frac{k_z}{k_0}.
\label{eq:npeff}
\end{equation}
In terms of the total MP momentum in Eq.~\eqref{eq:ocdmomentum} and the field energy
in Eq.~\eqref{eq:ocdenergy}, the effective phase refractive index can also be written as
$ n_\mathrm{p,eff}=c|\mathbf{P}_\mathrm{MP}|/E_\mathrm{field}$.
This relation is justified by the numerical OCD simulations of Sec.~\ref{sec:simulations}.
From the characteristic equation, one also obtains the effective group refractive index $n_\mathrm{g,eff}$,
given by
\begin{equation}
 n_\mathrm{g,eff}=\frac{\partial k_z}{\partial k_0}.
\label{eq:ngeff}
\end{equation}
There also exists an alternative well-known expression for the group refractive index
as a ratio of the integrals of the energy density and the Poynting vector.
Using Eqs.~\eqref{eq:ocdenergy} and \eqref{eq:ocdmomentum}, we then obtain
$ n_\mathrm{g,eff}=E_\mathrm{field}/(c|\mathbf{P}_\mathrm{field}|)$.


\subsection{Field modes and cutoff conditions}

\subsubsection{TE and TM modes}

First, we consider the special cases of the $TE$ and $TM$ modes.
For the $TE$ modes, we have  $E_z=0$, from which it follows that $A=C=0$.
In this case, the left hand sides of Eqs.~\eqref{eq:bc3} and \eqref{eq:bc4}
become zero. Equation \eqref{eq:bc4} is then satisfied if
$l=0$ and Eq.~\eqref{eq:bc3} leads to
the characteristic equation for the $TE$ modes given by
\begin{equation}
 \frac{J_1(hR)}{hRJ_0(hR)}=-\frac{K_1(qR)}{qRK_0(qR)},
 \label{eq:characteristicTE}
\end{equation}
where we have used the Bessel function relations $J_0'(x)=-J_1(x)$ and $K_0'(x)=-K_1(x)$.
The remaining Eq.~\eqref{eq:bc1} is trivially satisfied and Eq.~\eqref{eq:bc2} relates the
nonzero field amplitudes $B$ and $D$ to each other.

Respectively, for the $TM$ modes we have $H_z=0$, from which it follows that $B=D=0$.
In this case, the right hand sides of Eqs.~\eqref{eq:bc3} and \eqref{eq:bc4} become zero.
 Equation \eqref{eq:bc3} is then satisfied if
$l=0$ and Eq.~\eqref{eq:bc4} leads to
the characteristic equation for the $TM$ modes given by
\begin{equation}
 \frac{J_1(hR)}{hRJ_0(hR)}=-\frac{n_2^2K_1(qR)}{n_1^2qRK_0(qR)},
 \label{eq:characteristicTM}
\end{equation}
where we have again used the Bessel function relations $J_0'(x)=-J_1(x)$ and $K_0'(x)=-K_1(x)$.
The remaining Eq.~\eqref{eq:bc2} is trivially satisfied and Eq.~\eqref{eq:bc1} relates the
nonzero field amplitudes $A$ and $C$ to each other.


The $TE_{0,m}$ and $TM_{0,m}$ modes are numbered so that
the lowest mode number $m$ corresponds to the solution of the characteristic equation
for which the value of $hR$ is the smallest. The cutoff condition
for the existence of the $TE_{0,m}$ and $TM_{0,m}$ modes is given by \cite{Yariv2007}
\begin{equation}
 \Big(\frac{R}{\lambda_0}\Big)_{0,m}\ge\frac{x_{0,m}}{2\pi\sqrt{n_1^2-n_2^2}},
\end{equation}
where $x_{0,m}$, $m\ge1$, is the $m$th zero of the Bessel function $J_0(x)$. The first three zeros
of the Bessel function $J_0(x)$ are $x_{0,1}=2.405$, $x_{0,2}=5.520$,
and $x_{0,3}=8.654$.

\subsubsection{EH and HE modes}

Second, we consider the general solution of the characteristic equation
in Eq.~\eqref{eq:characteristic}. Solving the characteristic equation
for $J_l'(hR)/(hRJ_l(hR))$ gives
\begin{align}
&\frac{J_l'(hR)}{hRJ_l(hR)}\nonumber\\
&=-\Big(\frac{n_1^2+n_2^2}{2n_1^2}\Big)\frac{K_l'(qR)}{qRK_l(qR)}
    \pm\Big[\Big(\frac{n_1^2-n_2^2}{2n_1^2}\Big)^2\Big(\frac{K_l'(qR)}{qRK_l(qR)}\Big)^2\nonumber\\
&\hspace{0.5cm}+\Big(\frac{lk_z}{n_1k_0}\Big)^2\Big(\frac{1}{(qR)^2}+\frac{1}{(hR)^2}\Big)^2\Big]^{1/2}.
\end{align}
Here, the plus sign corresponds to the $EH$ modes
and the minus sign corresponds to the $HE$ modes.
The cutoff conditions
for the existence of the $EH_{1,m}$ and $HE_{1,m}$ modes are given by \cite{Yariv2007}
\begin{equation}
 \Big(\frac{R}{\lambda_0}\Big)_{1,m}^{EH}\ge\frac{x_{1,m}}{2\pi\sqrt{n_1^2-n_2^2}},\hspace{0.3cm}
 \Big(\frac{R}{\lambda_0}\Big)_{1,m}^{HE}\ge\frac{x_{1,m-1}}{2\pi\sqrt{n_1^2-n_2^2}},
\end{equation}
where $x_{1,m}$, $m\ge1$, is the $m$th zero of the Bessel function $J_1(x)$ and $x_{1,0}=0$.
The first three zeros of the Bessel function $J_1(x)$ are $x_{1,1}=3.832$, $x_{1,2}=7.016$,
and $x_{1,3}=10.173$. The cutoff values of $(R/\lambda_0)_{1,m}$ for the $HE_{1,m}$ modes are smaller
than for the $EH_{1,m}$ modes, and especially, the mode $HE_{1,1}$ does not have a cutoff.
The cutoff conditions
for the existence of the $EH_{l,m}$ and $HE_{l,m}$ modes with $l\ge2$ are given by \cite{Yariv2007}
\begin{equation}
 \Big(\frac{R}{\lambda_0}\Big)_{l,m}^{EH}\ge\frac{x_{l,m}}{2\pi\sqrt{n_1^2-n_2^2}},\hspace{0.3cm}
 \Big(\frac{R}{\lambda_0}\Big)_{l,m}^{HE}\ge\frac{z_{l,m}}{2\pi\sqrt{n_1^2-n_2^2}},
\end{equation}
where $x_{l,m}$, $m\ge1$, is the $m$th zero of the Bessel function $J_l(x)$ and $z_{l,m}$
is the $m$th root of equation $zJ_l(z)=(l-1)(1+n_1^2/n_2^2)J_{l-1}(z)$.

\subsection{Angular momentum and relation to Laguerre-Gaussian modes}

The Bessel modes studied in this work are eigenmodes
of step-index circular waveguides as discussed above.
The step-index profile is used in most single-mode fibers
and in some multi-mode fibers.
In counter distinction, Laguerre-Gaussian modes are eigenmodes of graded-index circular
waveguides with a parabolic refractive index profile.
The parabolic profile is the most common profile in graded-index multi-mode fibers as
it minimizes modal dispersion by continual refocusing of the rays in the fiber core.
The Bessel and Laguerre-Gaussian beams can both
carry angular momentum. This has been experimentally demonstrated by
rotating microparticles in optical tweezers
\cite{Grier2003,He1995,Friese1996,Simpson1997,Friese1998,ONeil2002,CarcesChavez2003,Adachi2007}.

\section{\label{sec:simulations}Simulations}

Next, we apply the OCD model to illustrate the node structure of the
atomic MDW and the actual atomic displacements
due to a light pulse in a circular waveguide made of fused silica.
The light pulse of Eqs.~\eqref{eq:efield1}--\eqref{eq:hfield2}
is assumed to have a total electromagnetic energy of $U_0=1$ $\mu$J
and a vacuum wavelength of $\lambda_0=1550$ nm.
The corresponding photon number is $N_\mathrm{ph}=E_\mathrm{field}/\hbar\omega_0=7.803\times 10^{12}$.
In our examples, the waveguide core radius is $R=\lambda_0/2$.
The relative spectral width of the pulse used in the simulations
is $\Delta\omega/\omega_0=\Delta k_0/k_0=0.05$,
which corresponds to the temporal full width at half maximum (FWHM)
of $\Delta t_\mathrm{FWHM}=27$ fs.
The FWHM is fixed to this small value, which is close to the engineering feasibility limit,
to make the node structure of the MDW visible
in the same scale with the longitudinal Gaussian envelope of the pulse.
In our simulations, we use temporal discretization of $h_t=2\pi/(40\omega_0)$
and spatial discretization of
$h_z=\lambda/40$, where $\lambda=\lambda_0/n_\mathrm{p,eff}$
is the effective wavelength of the studied mode in the waveguide.
This discretization is sufficiently dense compared to the scale of the harmonic cycle
of the optical field.
The flowchart of the OCD simulation is described in
Appendix C of \cite{Partanen2017c}.

The simulations are performed for a waveguide whose core is
assumed to be made of fused silica,
which is the standard material used in optical fibers.
The cladding layer is assumed to be vacuum meaning that there is no cladding.
The corresponding fields are a special case of Eqs.~\eqref{eq:efield1}--\eqref{eq:hfield2}
obtained by setting $n_2=1$.
The phase and group refractive indices of fused silica are given by
$n_\mathrm{p}=1.4440$, and $n_\mathrm{g}=1.4626$
for $\lambda_0=1550$ nm \cite{Malitson1965}. As the material dispersion is not very large,
in this work, we neglect the material dispersion and use the phase refractive index only.
However, the material dispersion could be accounted for as explained in \cite{Partanen2017e}.
The waveguide dispersion is here the dominating form of dispersion and it is
described by the effective phase and group refractive indices of the waveguide
geometry following from Eqs.~\eqref{eq:npeff} and \eqref{eq:ngeff}.
For the $TE_{0,1}$ mode studied below, the effective phase refractive index of our
waveguide geometry is $n_\mathrm{p,eff}=1.1330$
and the effective group refractive index is $n_\mathrm{g,eff}=1.5856$.
Respectively, for the $HE_{1,1}$ mode that is studied in the second example,
the effective phase refractive index is $n_\mathrm{p,eff}=1.3033$
and the effective group refractive index is $n_\mathrm{g,eff}=1.5355$.
The mass density of fused silica used in the OCD simulations is
$\rho_0=2200$ kg/m$^3$ \cite{Bruckner1970}, and
the Lam\'e elastic constants are $\lambda_\mathrm{L}=15.4$ GPa
and $\mu_\mathrm{L}=31.3$ GPa \cite{DeJong2000}.

\subsection{Simulations of the atomic mass density wave}

\begin{figure*}
\centering
 \includegraphics[width=0.85\textwidth]{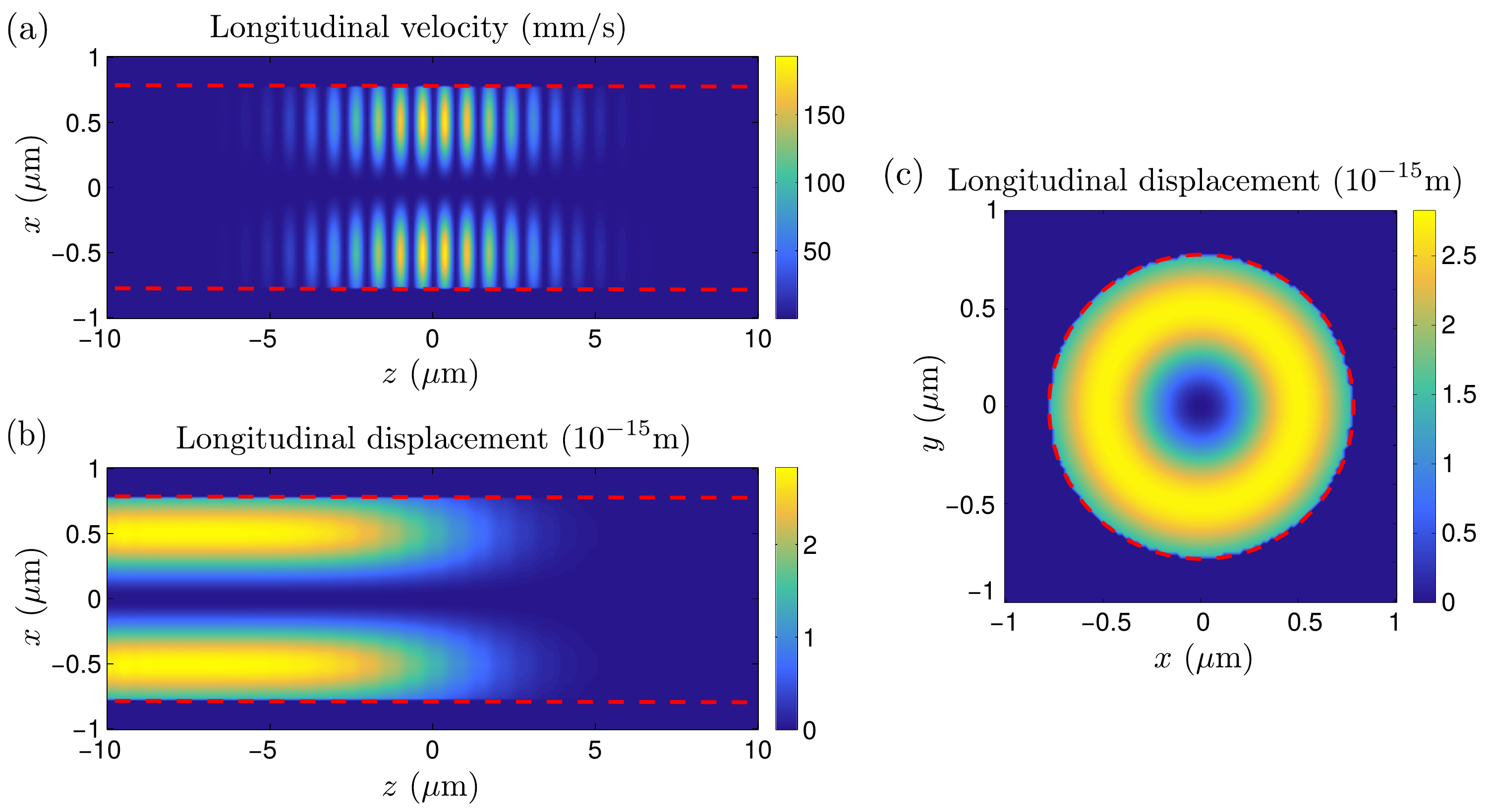}
\caption{\label{fig:TE01}
Simulation of the mass transfer due to a temporally Gaussian $TE_{0,1}$ mode light
pulse in a silica fiber (see also Visualization 1). (a) Simulated longitudinal atomic velocity of the MDW in
the fiber as a function of position in the plane $y=0$ $\mu$m.
(b) Longitudinal atomic displacement in the fiber as a function of position in the plane $y=0$ $\mu$m.
(c) Longitudinal atomic displacement in the fiber cross section $z=0$ $\mu$m just after the pulse has gone.
The pulse energy is $E_\mathrm{field}=1$ $\mu$J, the wavelength is $\lambda_0=1550$ nm,
and the duration is $\Delta t_\mathrm{FWHM}=27$ fs.
The radius of the circular fiber core is $R=\lambda_0/2$
and it is surrounded by vacuum.
The dashed line shows the boundaries of the fiber.}
\end{figure*}

First, we study the atomic MDW of a temporally Gaussian $TE_{0,1}$ light pulse for which $l=0$.
See Visualization 1 for a video file of the simulation.
Figure \ref{fig:TE01}(a) shows the longitudinal atomic velocities of the simulated
MDW as a function of position in the plane $y=0$ $\mu$m when the $TE_{0,1}$ pulse center
is propagating at the position $z=0$ $\mu$m.
In vacuum, outside the waveguide core at $x<-\lambda_0/2$ or at $x>\lambda_0/2$,
the atomic velocities are zero as the atomic MDW does not exist.
Inside the waveguide core, the atomic velocities in the MDW
follow the field intensity. The node structure of the field is then also well seen
in the atomic MDW. The atomic velocities are zero at the center of the waveguide
at the position $x=0$ $\mu$m following the $TE_{0,1}$ mode profile. 
When the Gaussian envelope of the pulse drops to zero
on the left of $z=-5$ $\mu$m and on the right of $z=5$ $\mu$m,
also the atomic velocities of the MDW become zero.

Figure \ref{fig:TE01}(b) shows the longitudinal atomic displacements of the simulated MDW as a function of
position along $z$ axis in the plane $y=0$ $\mu$m when the $TE_{0,1}$ pulse center
is propagating at the position $z=0$ $\mu$m corresponding to the atomic velocities in Fig.~\ref{fig:TE01}(a).
One can see that the atomic displacements are largest at lateral positions of maximum
field intensity. At positions $z>5$ $\mu$m, the atomic displacement is zero
as the light pulse has not yet reached these positions, and therefore, these atoms
have not yet experienced the optical force density.
In vacuum, outside the waveguide core, the atomic displacement in Fig.~\ref{fig:TE01}(b) is zero
as the atomic velocities in Fig.~\ref{fig:TE01}(a).
Behind the light pulse at $z<-5$ $\mu$m, the atomic displacement
has obtained a value that is in the femtosecond time scale
practically constant in time, but which varies in the lateral direction.
This atomic displacement is also the origin of the transferred
mass of the MDW and the corresponding mass energy term in Eq.~\eqref{eq:ocdenergy}
since it is related to the denser spacing of atoms at the position of the light pulse.

The longitudinal atomic displacement in the waveguide cross section
just after the light pulse has gone is presented in Fig.~\ref{fig:TE01}(c).
These atomic displacements correspond to the atomic displacements
behind the light pulse in Fig.~\ref{fig:TE01}(b).
The atomic displacement averaged over the waveguide cross section
is given by $\Delta r_\mathrm{MDW}=\delta M/(\rho_0\pi R^2)=2.135\times 10^{-15}$ m.
The atomic displacements that the light pulse leaves behind
are relaxed to equilibrium only in the much larger time scale of elastic waves.
The relaxation dynamics of the waveguide in the longer time scale is studied in
more detail in Sec.~\ref{sec:relaxation}.

The OCD simulation results in Fig.~\ref{fig:TE01} are fully consistent with the
MP quasiparticle model of \cite{Partanen2017c,Partanen2017e}
as the total transferred mass and momentum following from the atomic velocities and displacements
equal the quasiparticle model results within the 6-digit numerical accuracy of the simulation.
This very good accuracy is related to the monochromatic field approximation
of light pulses used in the OCD simulations. It is not exact for our short pulses,
but the relative error is found to be less than 1\%, which is sufficient for our visualization purposes.
Therefore, our results must be interpreted to indicate that full agreement between the OCD and MP quasiparticle
models is obtained in the limit of long narrow-band pulses in which limit our pulse approximation becomes exact.
For the total transferred mass, the MP quasiparticle model gives an expression
$\delta M=(n_\mathrm{p,eff}n_\mathrm{g,eff}-1)N_\mathrm{ph}\hbar\omega_0/c^2$
that is obtained from the homogeneous field result presented in \cite{Partanen2017e}
by simply replacing the phase and group refractive indices of the material
with the conventional effective values for the studied mode in the waveguide geometry.
The numerical OCD model gives the total transferred mass for our  $TE_{0,1}$ mode light pulse as
$\delta M=\int[\rho_\mathrm{a}(\mathbf{r},t)-\rho_0]d^3r=8.862\times 10^{-24}$ kg,
which corresponds to the value obtained by using the MP quasiparticle model formula.
For the momentum of the MDW, the OCD simulations give
$p_\mathrm{MDW}=|\int\rho_\mathrm{a}(\mathbf{r},t)\mathbf{v}_\mathrm{a}(\mathbf{r},t)d^3r|=1.676\times 10^{-15}$ kgm/s,
which corresponds to the MP quasiparticle model value obtained by using
$p_\mathrm{MDW}=(n_\mathrm{g,eff}-1/n_\mathrm{g,eff})N_\mathrm{ph}\hbar\omega_0/c$.

\begin{figure*}
\centering
 \includegraphics[width=\textwidth]{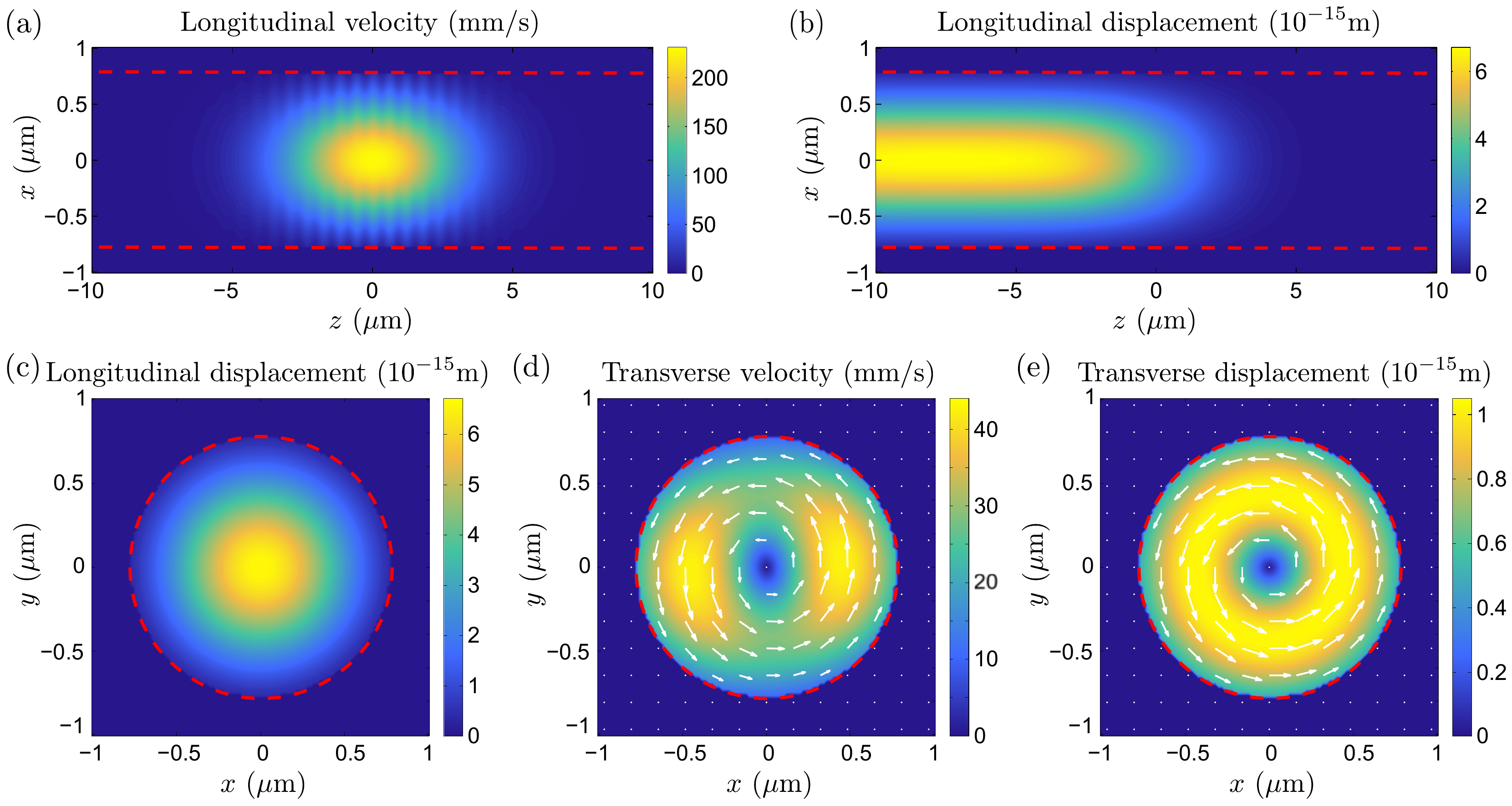}
\caption{\label{fig:HE11}
Simulation of the mass transfer due to a temporally Gaussian $HE_{1,1}$ mode light
pulse in a silica fiber (see also Visualization 2). (a) Longitudinal atomic velocity of the MDW
in the fiber as a function of position in the plane $y=0$ $\mu$m.
(b) Longitudinal atomic displacement in the fiber as a function of position in the plane $y=0$ $\mu$m.
(c) Longitudinal atomic displacement in the fiber cross section $z=0$ $\mu$m just after the pulse has gone.
(d) Instantaneous transverse atomic velocity of the MDW
in the fiber as a function of position in the fiber cross section
at the position of the pulse center at $z=0$ $\mu$m.
(e) Transverse atomic displacement in the fiber as a function of position in the fiber
cross section $z=0$ $\mu$m just after the light pulse has gone.
The pulse energy is $E_\mathrm{field}=1$ $\mu$J, the wavelength is $\lambda_0=1550$ nm,
and the duration is $\Delta t_\mathrm{FWHM}=27$ fs.
The radius of the circular fiber core is $R=\lambda_0/2$
and it is surrounded by vacuum.
The dashed line shows the boundaries of the fiber.}
\end{figure*}

Another interesting case is the $HE_{1,1}$ mode light pulse for which $l=1$.
This pulse does not have a cutoff, and in contrast to the $TE_{0,1}$ pulse above,
it carries angular momentum of $\hbar$ per photon.
See Visualization 2 for a video file of the simulation.
Figure \ref{fig:HE11}(a) depicts the atomic velocities of the simulated
MDW as a function of position in the plane $y=0$ $\mu$m when the $HE_{1,1}$ pulse center
is propagating at the position $z=0$ $\mu$m.
Inside the waveguide core, the atomic velocities in the MDW again
follow the field intensity. As the $HE_{1,1}$ mode is not a transverse mode,
the pulse does not have as clear peaks and throughs as the $TE_{0,1}$ pulse above.
Also, in contrast to the $TE_{0,1}$ pulse, the atomic velocities obtain their
maximum values at the center of the waveguide at the position $x=0$ $\mu$m
in accordance with the $HE_{1,1}$ mode profile.
On the left and right of the light pulse, the atomic velocities in the MDW
are again zero as expected.

Figure \ref{fig:HE11}(b) depicts the longitudinal atomic displacements of the simulated MDW as a function of
position along $z$ axis in the plane $y=0$ $\mu$m when the $HE_{1,1}$ pulse center
is propagating at the position $z=0$ $\mu$m. These atomic displacements correspond
to the atomic velocities in Fig.~\ref{fig:HE11}(a).
One can again see that the longitudinal atomic displacements are largest at lateral positions of maximum
field intensity. In front of the light pulse at positions $z>5$ $\mu$m, the atomic displacement is
again zero and, behind the light pulse at $z<-5$ $\mu$m, the atomic displacement
has obtained a value that is temporally approximatively constant in the femtosecond time scale.
The longitudinal atomic displacements of the $HE_{1,1}$ pulse
just after the light pulse has gone is presented in the waveguide cross section in Fig.~\ref{fig:HE11}(c).
These atomic displacements correspond to the atomic displacements
after the light pulse in Fig.~\ref{fig:HE11}(b).
The atomic displacement averaged over the waveguide cross section
is given by $\Delta r_\mathrm{MDW}=\delta M/(\rho_0\pi R^2)=2.683\times 10^{-15}$ m.

The simulation results are again in full correspondence with the MP quasiparticle
model of \cite{Partanen2017c} and \cite{Partanen2017e}.
Again, note that full agreement is obtained by using the monochromatic
field approximation of light pulses in the OCD model.
For the total transferred mass of our  $HE_{1,1}$ pulse, we obtain
$\delta M=1.114\times 10^{-23}$ kg and, for the momentum of the MDW,
we obtain a value $p_\mathrm{MDW}=2.175\times 10^{-15}$ kgm/s.
These values are slightly larger than the values in the case of
the $TE_{0,1}$ pulse due to the smaller evanescent field that propagates in vacuum
outside the fiber core and that does not directly contribute to the atomic MDW. 
In the MP quasiparticle model, this difference between the quantities
is accounted for by the effective phase and group refractive indices of the waveguide
corresponding to the somewhat smaller waveguide dispersion in the $HE_{1,1}$ case.

Figure \ref{fig:HE11}(d) presents the transverse components of the atomic velocities
in the MDW at a particular instance of time in the middle of the $HE_{1,1}$ light pulse.
The transverse components of the atomic velocities follow from the nonzero transverse components
of the optical force density and the Poynting vector, which do not exist in the case of the $TE_{0,1}$ pulse above.
As described in \cite{Partanen2018a}, the transverse components of the atomic
velocities are related to the transfer of a substantial part of the total angular
momentum of light in a medium.
The angular momentum of the electromagnetic field of the $HE_{1,1}$ pulse
obtained using the OCD model is
$\mathbf{J}_\mathrm{field}=0.586N_\mathrm{ph}\hbar\hat{\mathbf{z}}$
and the angular momentum of the MDW is
$\mathbf{J}_\mathrm{MDW}=0.414N_\mathrm{ph}\hbar\hat{\mathbf{z}}$.
The sum of these angular momentum contributions is the total angular momentum
of the coupled state of the field and matter, given by
$\mathbf{J}_\mathrm{MP}=N_\mathrm{ph}\hbar\hat{\mathbf{z}}$,
which is essentially an integer multiple of $\hbar$.
This result verifies the coupling of the field and medium
components of the total angular momentum of a light pulse
in the case of a waveguide geometry.
Previously, this coupling has been studied in the case of a homogeneous
medium in \cite{Partanen2018a}.

The cross section of the total transverse atomic displacement field that the
$HE_{1,1}$ pulse leaves behind, and which follows from
the transverse atomic velocities in Fig.~\ref{fig:HE11}(d), is
depicted in Fig.~\ref{fig:HE11}(e). Together with the longitudinal
atomic displacements in Fig.~\ref{fig:HE11}, these transverse atomic displacements
are again relaxed to equilibrium in the time scale of elastic waves.

\subsection{\label{sec:relaxation}Simulations of the elastic relaxation dynamics}

In the OCD simulations, the optical force density of the light pulse
displaces atoms from their initial equilibrium positions by different amounts
in the cross section of the fiber.
Therefore, there will be strain and related elastic forces acting between atoms
after the light pulse has gone and when, accordingly, the optical force is zero.
Thus, the after-the-light-pulse elastic relaxation dynamics is a novel feature
of the OCD model that results from rejecting the approximation of fixed atoms.
In this work, we simulate the elastic relaxation of the medium by using
elasticity theory without including terms that lead to thermalization.
The elastic relaxation dynamics
has been previously simulated in the case of a homogeneous
material block in \cite{Partanen2017c} and it has been qualitatively
described in the case of optical fibers in \cite{Partanen2017e}.

Here, we simulate the beginning of the relaxation of the nonequilibrium atomic
displacements of the waveguide
after the $TE_{0,1}$ light pulse of Fig.~\ref{fig:TE01} has gone.
The initial atomic displacement for this relaxation process is
that presented in Fig.~\ref{fig:TE01}(c). 
We assume that the fiber is infinitely long, which is realized in the simulations
by setting periodic boundary conditions in the longitudinal direction just after
the light pulse has gone. The damping of the
elastic waves in a closed system is not accounted for by the elastic force density
in Eq.~\eqref{eq:elasticforcedensity}. Therefore, the elastic waves
that we obtain do not attenuate. Hence, we cannot
simulate how the equilibrium of atomic positions is obtained at the end of the relaxation.
Instead, we focus on the first
nanosecond when the damping of the elastic waves can be neglected.

\begin{figure*}
\centering
 \includegraphics[width=\textwidth]{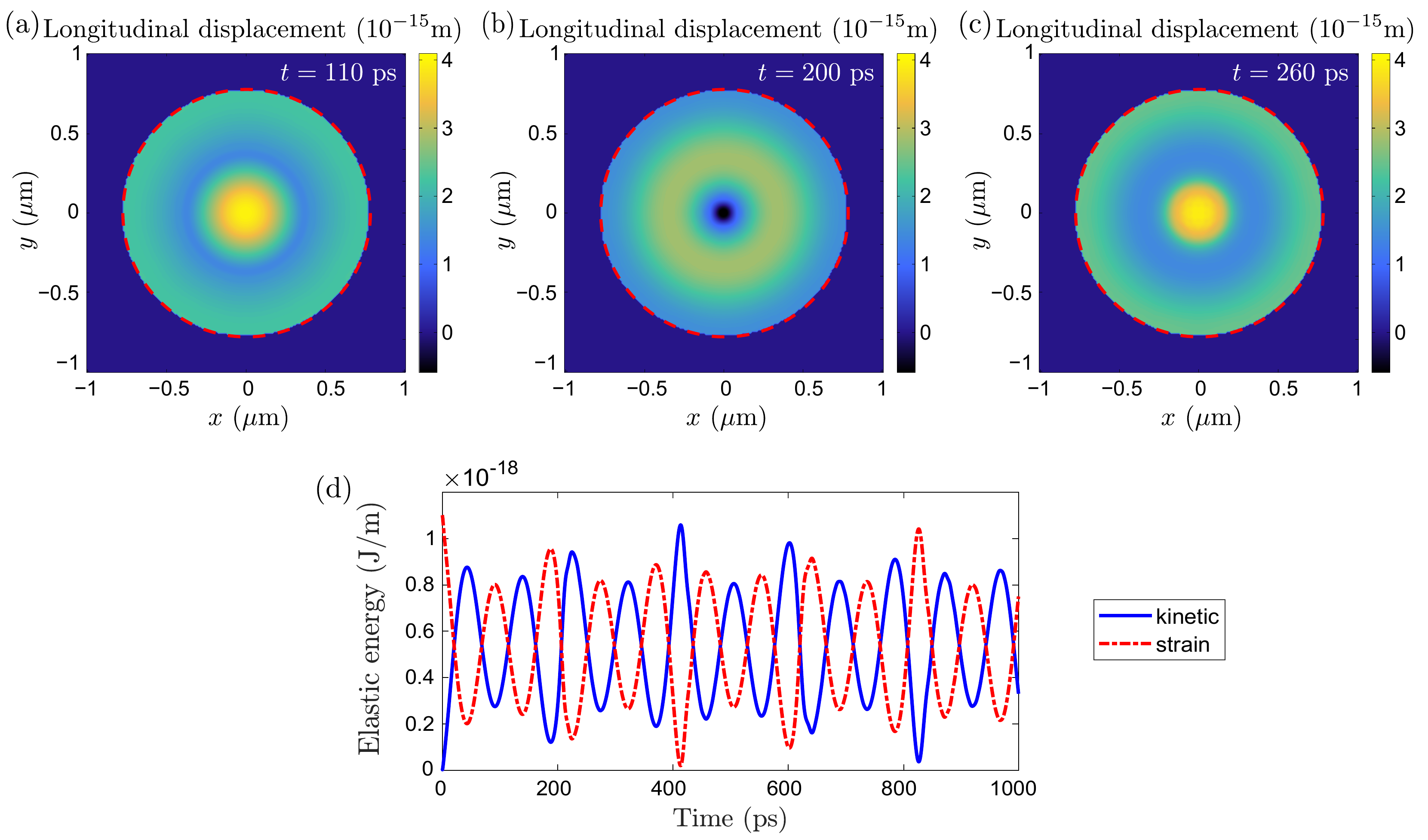}
\caption{\label{fig:relaxation}
Simulation of the elastic relaxation dynamics of the silica waveguide after the $TE_{0,1}$ mode light
pulse (see also Visualization 3). The pulse energy is $E_\mathrm{field}=1$ $\mu$J, the wavelength is $\lambda_0=1550$ nm, and $\Delta t_\mathrm{FWHM}=27$ fs.
Longitudinal atomic velocity of the MDW at $z=0$ $\mu$m
cross section is shown at (a) $t=110$ ps,
(b) $t=200$ ps, and (c) $t=260$ ps after the pulse has gone.
The radius of the circular fiber core is $R=\lambda_0/2$
and it is surrounded by vacuum.
The dashed line shows the fiber boundary.
(d) The alternation of the
elastic (kinetic and strain) energies per unit length of the fiber as a function of time.
The elastic waves are gradually damped and thermalized
by absorption and scattering not included in our simulations.
Due to the periodic boundary conditions used in the simulation
of the elastic relaxation, the interface effects at the ends of the
fiber are not present.}
\end{figure*}

Figure \ref{fig:relaxation} presents the relaxation dynamics of the waveguide
after the $TE_{0,1}$ light pulse of Fig.~\ref{fig:TE01} has gone. It shows the longitudinal
atomic displacement in the waveguide cross section at three instances of time in the picosecond time scale.
See Visualization 3 for a video file of the simulation.
Figure \ref{fig:relaxation}(a) shows the longitudinal atomic displacement at $t=110$ ps,
Fig.~\ref{fig:relaxation}(b) at $t=200$ ps, and Fig.~\ref{fig:relaxation}(c) at $t=260$ ps.
One can see that the minima and maxima of the atomic displacement switch their positions in the course
of time. In Figs.~\ref{fig:relaxation}(a) and \ref{fig:relaxation}(c), the maximum atomic
displacement is located at the origin while, in Fig.~\ref{fig:relaxation}(b), at the origin,
the atomic displacement obtains its minimum value.
This can be understood as elastic shear waves that propagate from the circular fiber boundary
to the center and back. The atomic displacement minima and maxima alternate,
but their time-evolution is not fully periodic since the initial
state of the relaxation process is not an eigenstate of the elastic oscillation
of the waveguide.

Figure \ref{fig:relaxation}(d) shows the kinetic and strain energies of the
elastic relaxation waves per unit length of the fiber as a function of time. At $t=0$,
the elastic energy is pure strain energy as the light pulse
has only displaced atoms from their initial equilibrium positions but does not
have left any net kinetic energy to them. At $t>0$, the kinetic and strain energies
alternate as a function of time so that their sum remains constant
due to the conservation of energy.
The highest peaks and throughs in the kinetic and strain energies at
$t=413$ ps and $t=826$ ps are seen to be located close to times
at which the shear waves with velocity $v_\perp=\sqrt{\mu_\mathrm{L}/\rho_0}=3770$ m/s
have propagated distances that are multiples of the waveguide diameter.

The value of the total elastic energy, left in the waveguide
after the light pulse has gone, can be used to estimate
the effective imaginary part of the refractive index $n_\mathrm{i}$, which
corresponds to the dissipation of the field energy when the field propagates in the waveguide.
The total elastic energy left in the waveguide per unit length is $1.1\times 10^{-18}$ J/m
as can be read from Fig.~\ref{fig:relaxation}(d).
Dividing this dissipated energy by the field energy $U_0$ of the pulse, gives
the attenuation coefficient $1.1\times 10^{-12}$ 1/m. When this value
is compared with the conventional formula for the attenuation coefficient,
given by $\alpha=2n_\mathrm{i}k_0$, we obtain $n_\mathrm{i}=1.4\times 10^{-19}$.
This is vastly smaller than the imaginary part of the refractive index
due to other physical nonidealities in highly transparent optical waveguides
that are not accounted for in the present simulations.

All of the elastic strain energy that is left in the crystal after the light pulse
has gone will be converted into heat in the course of time.
The detailed simulation of the thermalization of strain is beyond
the scope of the present manuscript.
The dissipation of field energy into energy of elastic waves
is an additional very small loss mechanism to the dominant
losses in realistic transparent materials.
We find the MDW-related dissipation effect even though
we initially assume that the refractive index is real
valued and the field is not damped. In self-consistent calculations, the corresponding
damping of the field could be accounted for through
the self-consistently determined value of the imaginary
part of the refractive index. Note, however, that the self-consistent calculations accounting for the imaginary
part of the refractive index coming from the elastic MDW losses alone
are not meaningful as this extremely small imaginary part refers
to damping that is several orders of magnitude smaller than
the total damping in the best transparent solids.

Since the velocity of light is vastly larger than the velocity of sound,
the relaxation process practically starts only after our short light pulse has gone.
In this approximation, during the relaxation process, the total external force on the medium atoms
is zero. Therefore, in accordance with Newton's first law, the center
of mass of the waveguide remains at rest during the relaxation process since the light pulse
has only displaced atoms from their initial equilibrium positions but does not
have left any net kinetic energy to them as discussed above.
This contrasts to the state of motion of the waveguide under the influence of the optical force
of the light pulse, which moves the center of mass of the waveguide
slightly forwards by the atomic MDW that is present in the waveguide while the light pulse
is propagating in it.

The same kind of relaxation waves, as obtained
for the longitudinal atomic displacements in Fig.~\ref{fig:relaxation},
are also obtained for the transverse atomic displacements in the relaxation of
the atomic displacements due to optical modes
carrying angular momentum. These waves are also elastic
shear waves in nature. Compressional waves are not obtained
as the light pulse does not leave any mass density differences
behind in the middle of the waveguide. As shown in \cite{Partanen2017c},
compressional waves are found in the case that a light pulse crosses
an interface between materials of different refractive indices.
Then, the optical surface force displaces atoms at the surface
and these displacements are relaxed in terms of both compressional
and shear waves. In the case of optical fibers,
compressional waves can originate from
the processes at the ends of the fiber or at any positions
where there are changes in the refractive index.

\section{\label{sec:comparison}Comparison of the OCD and MP quasiparticle models}

The correspondence between the OCD and MP quasiparticle model results
is summarized in Table \ref{tbl:table}. 
Full agreement is obtained within the numerical accuracy of the simulations and
the monochromatic field approximation of light pulses used in the OCD model.
The results for the energy and momentum
of the MP and its field and medium parts are identical to the
results for a homogeneous medium in \cite{Partanen2017c}
with an exception that we must now use the effective phase and group
refractive indices of the waveguide in the MP quasiparticle model.
Consequently, we obtain the ratio
$n_\mathrm{p,eff}n_\mathrm{g,eff}-1$ for the MDW's and the
electromagnetic field's shares
of energies and momenta. This result is directly related to the
Lorentz transformation and the covariance principle of the special
theory of relativity in the same way as described in the case
of a homogeneous medium in \cite{Partanen2017c}.

Also, note that many previous works, which are not directly related
to the MP theory of light, obtain the Minkowski momentum
as the total momentum of light \cite{Barnett2010b,Bliokh2017a,Bliokh2017b,Leonhardt2014,Brevik2017}
in agreement with our result in Table \ref{tbl:table}.
Since these works are typically based on the approximation of fixed atoms,
they are not able to unambiguously separate the total momentum of light between
the field and the atoms. In the work of Leonhardt \cite{Leonhardt2014},
the atoms are allowed to move, but the medium is assumed to be incompressible,
which effectively makes one to neglect the coupled field-medium dynamics
studied in the present work.

\begin{table}
 \centering
 \renewcommand{\arraystretch}{2.0}
\footnotesize
 \caption{\label{tbl:table}
 The transferred mass, the total momentum, the field's share of the
 momentum, and the MDW's share of the momentum calculated
 by using the MP and OCD models. The MP column shows the per photon
value obtained by dividing the total quantities of the MP model with the photon number
of the pulse $N_\mathrm{ph}=E_\mathrm{field}/\hbar\omega_0$.}
\vspace{0.1cm}
\begin{tabular}{ccc}
   \hline
   & OCD & MP \\[4pt]
   \hline
   $\delta M$ & $\displaystyle\int\rho_\mathrm{MDW}d^3r$ & $\displaystyle(n_\mathrm{p,eff}n_\mathrm{g,eff}-1)\frac{\hbar\omega_0}{c^2}$ \\[4pt]
   $\mathbf{P}_\mathrm{MP}$       & $\displaystyle\int\Big(\rho_\mathrm{a}\mathbf{v}_\mathrm{a} + \frac{\mathbf{E}\times\mathbf{H}}{c^2}\Big)d^3r$ & $\displaystyle\frac{n_\mathrm{p,eff}\hbar\omega_0}{c}\hat{\mathbf{z}}$ \\[4pt]
   $\mathbf{P}_\mathrm{field}$      & $\displaystyle\int\frac{\mathbf{E}\times\mathbf{H}}{c^2}d^3r$ & $\displaystyle\frac{\hbar\omega_0}{n_\mathrm{g,eff}c}\hat{\mathbf{z}}$ \\[4pt]
   $\mathbf{P}_\text{MDW}$             & $\displaystyle\int\rho_\mathrm{a}\mathbf{v}_\mathrm{a}d^3r$ & $\displaystyle\Big(n_\mathrm{p,eff}-\frac{1}{n_\mathrm{g,eff}}\Big)\frac{\hbar\omega_0}{c}\hat{\mathbf{z}}$ \\[4pt]
   $\mathbf{J}_\mathrm{MP}$       & $\displaystyle\int\mathbf{r}\times\Big(\rho_\mathrm{a}\mathbf{v}_\mathrm{a} + \frac{\mathbf{E}\times\mathbf{H}}{c^2}\Big)d^3r$ & $l\hbar\hat{\mathbf{z}}$ \\[4pt]
   $\mathbf{J}_\mathrm{field}$       & $\displaystyle\int\mathbf{r}\times\Big(\frac{\mathbf{E}\times\mathbf{H}}{c^2}\Big)d^3r$ & $\displaystyle \zeta l\hbar\hat{\mathbf{z}}$ \\[4pt]
   $\mathbf{J}_\mathrm{MDW}$       & $\displaystyle\int\mathbf{r}\times\rho_\mathrm{a}\mathbf{v}_\mathrm{a}d^3r$ & $\displaystyle(1-\zeta)l\hbar\hat{\mathbf{z}}$ \\[4pt]
   \hline
 \end{tabular}
\end{table}

The waveguide dispersion also influences the separation of the total
angular momentum of light between the field and matter
as described by the parameter $\zeta=|\mathbf{J}_\mathrm{field}|/|\mathbf{J}_\mathrm{MP}|$
used in Table \ref{tbl:table}.
Analytic calculation using Eq.~\eqref{eq:ocdangularmomentum}
and the fields in Eqs.~\eqref{eq:efield1}--\eqref{eq:hfield2}
shows that the angular momenta of the field and the MDW are given by
\begin{align}
|\mathbf{J}_\mathrm{field}|
&=\frac{\pi^{3/2}k_0R^2}{c n_\mathrm{g,eff}\Delta k_0h^2}
\Big\{\frac{l}{2}\Big(\varepsilon_1|A|^2+\mu_0|B|^2\Big)\nonumber\\
&\hspace{0.4cm}\times\Big[J_l(Rh)^2-J_{l-1}(Rh)J_{l+1}(Rh)\Big]\nonumber\\
&\hspace{0.4cm}+\mathrm{sign}(l)\frac{n_\mathrm{p,eff}}{c}|AB|J_{l-1}(Rh)J_{l+1}(Rh)\Big\}\nonumber\\
&\hspace{0.4cm}+\frac{\pi^{3/2}k_0R^2}{c n_\mathrm{g,eff}\Delta k_0q^2}
\Big\{\frac{l}{2}\Big(\varepsilon_2|C|^2+\mu_0|D|^2\Big)\nonumber\\
&\hspace{0.4cm}\times\Big[K_l(Rq)^2-K_{l-1}(Rq)K_{l+1}(Rq)\Big]\nonumber\\
&\hspace{0.4cm}+\mathrm{sign}(l)\frac{n_\mathrm{p,eff}}{c}|CD|K_{l-1}(Rq)K_{l+1}(Rq)\Big\},
\end{align}
\begin{align}
|\mathbf{J}_\mathrm{MDW}|
&=\frac{(n_1^2-1)\pi^{3/2}k_0R^2}{c n_\mathrm{g,eff}\Delta k_0h^2}
\Big\{\frac{l}{2}\Big(\varepsilon_1|A|^2+\mu_0|B|^2\Big)\nonumber\\
&\hspace{0.4cm}\times\Big[J_l(Rh)^2-J_{l-1}(Rh)J_{l+1}(Rh)\Big]\nonumber\\
&\hspace{0.4cm}+\mathrm{sign}(l)\frac{n_\mathrm{p,eff}}{c}|AB|J_{l-1}(Rh)J_{l+1}(Rh)\Big\}\nonumber\\
&\hspace{0.4cm}+\frac{(n_2^2-1)\pi^{3/2}k_0R^2}{c n_\mathrm{g,eff}\Delta k_0q^2}
\Big\{\frac{l}{2}\Big(\varepsilon_2|C|^2+\mu_0|D|^2\Big)\nonumber\\
&\hspace{0.4cm}\times\Big[K_l(Rq)^2-K_{l-1}(Rq)K_{l+1}(Rq)\Big]\nonumber\\
&\hspace{0.4cm}+\mathrm{sign}(l)\frac{n_\mathrm{p,eff}}{c}|CD|K_{l-1}(Rq)K_{l+1}(Rq)\Big\}.
\end{align}
Therefore, using $|\mathbf{J}_\mathrm{MP}|=|\mathbf{J}_\mathrm{field}|+|\mathbf{J}_\mathrm{MDW}|$
one also obtains an analytic expression for the parameter
$\zeta=|\mathbf{J}_\mathrm{field}|/|\mathbf{J}_\mathrm{MP}|$.
In the case of a homogeneous medium with a refractive index $n$, we have
$\zeta=1/n^2$ as shown in \cite{Partanen2018a}.
In contrast, in the waveguide geometry studied in this work, we have not
found a simple expression for $\zeta$
in terms of the effective phase and group refractive indices of
the waveguide. One can argue that this is not surprising as the covariance
condition of the special theory of relativity used in the MP quasiparticle model relates the energies
and momenta of the field and the MDW to each other \cite{Partanen2017c} but
no such a simple quasiparticle model relation is known to exist for the separation of angular momentum.
In the case of angular momentum, only the total angular momentum of light,
which is a multiple of $\hbar$, is well known to be independent of the inertial reference frame.

Not using the monochromatic field approximation of light pulses in the OCD model
induces small changes in the relation between the OCD and MP quasiparticle models
for short pulses (e.g., the relative difference is $<1$\% for the pulses studied in Sec.~\ref{sec:simulations}).
This is related to the generally $k$-dependent factors of the
fields in Eqs.~\eqref{eq:efield1}--\eqref{eq:hfield2},
due to which the exact pulse shapes become slightly modified in comparison
with the shapes obtained by using the monochromatic field approximation.

\section{\label{sec:conclusions}Conclusions}
In conclusion, we have used the MP theory of light and the related
OCD model to simulate the atomic MDW of propagating light pulses
in a step-index circular waveguide geometry.
We found unambiguous correspondence between
the MP quasiparticle and OCD models in the description
of light propagation in waveguides.
Since part of the total energy and momentum of light propagates in vacuum
outside the waveguide core, the total magnitude of the atomic MDW is reduced.
This reduction takes place in
a way that is fully accounted for by using the well-known effective
phase and group refractive indices of the waveguide
in place of the corresponding indices of a homogeneous medium
in the MP quasiparticle theory for dispersive media formulated by us recently.

In this work, we have also studied the elastic
relaxation dynamics of the atomic displacements
that the light pulse has left in the waveguide. We have found that
these displacements are relaxed by elastic shear waves.
The OCD model allows detailed modeling
of the optoelastic dynamics of any material geometries
under the influence of the optical force field, e.g., one can simulate
the MDW in typical waveguides
used in engineering applications.
The OCD simulations can also be used for the optimization
of possible measurement setups planned for the experimental
verification of the atomic MDW associated with light.

After the present work was completed, the preprint in \cite{Picardi2018}
including results complementary to our work, came to our attention.
This preprint does not, however, separate the total angular momentum
of light into the field and the atomic MDW contributions as we do in
the present work. Therefore, the coupled
dynamical description of the electromagnetic field and the atoms
in the waveguide is not covered in the mentioned work.

%

\section*{Acknowledgments}
This work has been funded in part by the Academy of Finland under contract numbers 287074 and 318197.

\end{document}